\documentclass[12pt,epsfig]{article}
\usepackage{graphicx}
\usepackage{amssymb}
\newcommand{\beq}[1] {\begin{equation}\label{#1} }
\newcommand{\eeq} {\end{equation} }

\newcommand{\bea}[1]{\begin{eqnarray}\label{#1} }
\newcommand{\eea}{\end{eqnarray}}

\begin{document}

\vspace*{-0.5cm}
\begin{flushright}
OSU-HEP-02-10\\
FERMILAB-PUB-02/139-T
\end{flushright}
\vspace{0.5cm}

\begin{center}
{\Large
{\bf New Signal for Universal Extra Dimensions} }

\vspace*{1.5cm}
 C. Macesanu\footnote{Email address:  mcos@pas.rochester.edu}$^{,\dag,\ddag}$,
C.D. McMullen\footnote{Email address: mcmulle@okstate.edu}$^{,\dag}$
and S. Nandi\footnote{Email address: shaown@okstate.edu}$^{,\dag,\ddag}$

\vspace*{0.5cm}
$^{\dag}${\it Department of Physics, Oklahoma State University\\
Stillwater, Oklahoma, 74078\\}

$^{\ddag}${\it Fermi National Accelerator Laboratory\\
P.O. Box 500, Batavia, Il 60510}\footnote{Summer Visitor Program.}
\end{center}

\begin{abstract}
In the universal extra dimensions (UED) scenario, the tree level 
masses of the first 
level Kaluza-Klein (KK) excitations of Standard Model particles are essentially
degenerate. Radiative corrections will, however, lift this degeneracy, 
allowing the first level excitations to decay to the lightest KK
particle (LKP), which is the $\gamma^*$. KK number conservation implies
that the LKP is stable. Then, since the SM  particles radiated 
during these decays are rather soft, the observation of 
KK excitations production and decay   
in collider experiments will be quite difficult.
We propose to add to this model 
KK number violating interactions mediated by gravity, which allow
the $\gamma^*$ to decay to a photon and a KK graviton. For a variety a models 
and a large range of parameters, these decays will occur within the detector.
Thus, pair production of KK excitations will give rise to a 
striking collider signal,
consisting of two hard photons plus large missing energy 
(due to escaping gravitons). We evaluate the cross-section for these signals 
at the Tevatron and LHC, and derive the reach of these colliders in the
search for universal extra dimensions.
\end{abstract}

\section{Introduction}

In the past years there has been a resurgence of interest in the
phenomenological implications of models with extra dimensions. Such models
appear naturally in the context of string theories. However, since 
the
compactification scale is usually of order of the Planck mass, this
was thought to have few, if any, direct implications for experiments
at the current energy scale.
Recent developements \cite{witten}
have led to the construction of models where
the radius  of the  compact extra dimensions is of order 
inverse TeV, or even larger \cite{ADD}, with rich implications
for the phenomenology of present day colliders \cite{ed_pheno}. 


In this paper we are concerned with an extension of the basic model
proposed by Arkani-Hamed, Dimopoulos and Dvali (ADD) \cite{ADD}. In 
this universal extra dimensions (UED) scenario \cite{acd}, all the 
Standard Model (SM) fields, fermions as well as bosons, propagate 
in the bulk. This model has a number of interesting features, 
among which is the conservation of Kaluza-Klein
(KK) number for interactions involving
SM particles or their excitations. As a consequence,
KK excitations can be produced only in pairs at colliders; also,
tree-level corrections to SM processes are forbidden, making the
search for extra dimensions that much more difficult. Some
other features of this model are summarized in the following.

At tree level, the masses of the first level KK excitations are almost
degenerate. The splitting between the first level masses is due to the
SM mass terms, and is extremely small, except for particles
with large SM mass. Therefore, at tree level  most of the first level KK
excitations would be stable in the UED scenario,
due to KK number conservation. The
parameters of this model
may be subjected to restrictions due to cosmological constraints
on the existence of a large number of stable massive particles.
In order to bypass these restrictions,
some mechanisms provide for the decays of these excitations through KK-number
violating interactions mediated by gravity. Then, the experimental signature
for producing KK excitations at Tevatron or the Large Hadron
Collider (LHC) would be two jets plus
large missing energy. An analysis of the phenomenological implications 
of this model in hadron collider experiments has been performed in \cite{mmn}.
The results obtained indicate that, based on Tevatron Run I data, 
the compactification scale can be as low as 350 GeV. Moreover, the 
Tevatron Run II
can test for the existence of universal extra dimensions up to around
500 GeV, while the reach at LHC is about 3 TeV. 

Loop corrections, however, can give important contributions 
to the masses of the KK particles, thus potentially invalidating
the above analysis. For the UED model, these
corrections have been evaluated in \cite{cms1}. The result, dependent
on some assumptions concerning the renormalization of boundary terms, 
is that the radiative corrections to the tree level masses
 are typically of order 10\% for the strongly interacting particles 
(the heaviest one being the gluon) and of order few percent for the 
the leptons and electroweak gauge bosons (the photon being the lightest one).  
Naturally, the phenomenology of this model is quite different from the
case discussed above. The excitations of the SM quarks and gluons
produced at a hadron collider will cascade decay to the $\gamma^*$,
which is the lightest KK excitation. If this particle is stable, the 
experimental signature for this process will be the missing energy
carried away by the $\gamma^*$'s, and the soft SM particles radiated away
in the process of the cascade decays. The phenomenology 
of this model has been studied in \cite{cms2}. The total energy of these
particles will be of the order of the difference in mass between the 
KK particle which initiates the decay chain ($g^*$ or $q^*$) and the
$\gamma^*$, therefore they will be rather difficult to see in a hadron 
collider environment. Moreover, the transverse mising energy (which is the
quantity experimentally accesible) is also small, making the rejection
of SM backgrounds using this type of cuts quite difficult.  
The LHC reach for this model has been estimated in \cite{cms2} to be 
about 1.5 TeV.

The aim of this paper is to analyze the case when both types of decays,
gravity mediated and due to mass splitting, can occur. The experimental
consequences of this model can be quite remarkable. For example, in the case
when the decay widths of the first level KK excitations  due to mass splitting
are much larger than the gravity mediated decay widths, the gluon and
quark excitations produced at a hadron colider will cascade decay to
$\gamma^*$, which in turn will decay to a photon and a KK graviton. The
experimental signal in this case will be a striking two photon + missing
energy event. Moreover, since the photons are coming from the decay of a heavy
particle (the $\gamma^*$), their transverse momentum 
will be large, and the signal will be easy to separate from the 
SM background. The exact fraction of KK excitations decays  leading
to this kind of signal depends
on the parameters of the model, and will be discussed in the following
sections. 

The outline of this paper is as follows. In the next section we review the
essential features of the universal extra dimensions model at one loop. A 
mechanism for the gravity mediated decay of the first level KK excitations
is also presented.
In Section 3 we discuss the experimental consequences of the lightest KK
particle (LKP) $\gamma^*$ decaying into a photon and a graviton,  
and we give the collider reach for the discovery of UED at the
Tevatron and LHC in this scenario. We end with conclusions.

\section{Universal Extra Dimensions}

 The UED scenario is an extension of the SM in which all particles, fermions
as well as bosons, live in a $4+\delta$ dimensional brane, which is potentially
embedded in a larger space where only gravity propagates.
In this model, momentum conservation along the extra dimensions
dictates that the KK particles can be produced only in pairs. This would make
them much harder to see at present and future colliders; the limits
on the compactification scale of these extra dimensions are so far as low as
a few hundred GeV \cite{cms2,mmn,acd}.

While there are no {\it a priori} constraints on the number of universal
extra dimensions, we shall consider only the simpler case when there 
is only one such dimension. For $\delta = 1$, obtaining
the SM chiral fermions out of the zero modes of the 5-dimensional 
KK fields requires that the fifth dimension have an orbifold structure.  
For simplicity, this is usually taken to be $S_1/Z_2$. In this model, for
each SM Dirac fermion $q$ there are two 5-dimensional fermionic fields:
$q^{\bullet}$ and $q^{\circ}$. The first one is a doublet under SU(2), while
the second one is a singlet. Moreover, the left handed part of  
$q^{\bullet}$ and the right handed part of $q^{\circ}$ 
are taken to be even
under the $Z_2$ orbifold symmetry, while $q^{\bullet}_R$ and 
$q^{\circ}_L$ are taken to be odd
under the same symmetry, therefore projecting out half the zero modes
of each KK field. The remaining halves stand for the left and right-handed
parts of the SM chiral fermion: $ q_L = q^{\bullet}_0,\ q_R = q^{\circ}_0$.
The 4D interactions of the n$th$ 
KK excitation of the $q^{\bullet}$ and $q^{\circ}$
fields with the SM gauge bosons are given by:
$$
{\cal L}_{q_n - A} \ = \ - \bar{Q}^{\bullet}_n \left\{ Q e \not{X } + 
\frac{g_2}{\hbox{cos}\theta_W} \left( \frac{\tau_3}{2} 
- Q \hbox{sin}^2\theta_W \right) \not{Z }
\ + \ \frac{g_2}{\sqrt{2}}\left[ 
\begin{array}{cc}
0 & \not{W^+} \\
\not{W^-} & 0
\end{array}
\right]
\right\} Q^{\bullet}_n
$$ 
$$
- \  \bar{q}^{\circ}_n Q \left( e \not{X } - 
g_2 \frac{\hbox{sin}^2\theta_W}{\hbox{cos}\theta_W} \not{Z } \right)
q^{\circ}_n
$$
where $Q$ and $\tau_3$ are the charge and isospin of the corresponding
fermionic field, $X$ is the SM photon field, and $\theta_W, e$ and  $g_2$
are the SM Weinberg angle and the
electromagnetic and SU(2) four-dimensional coupling constants.

At tree level, the masses of the KK excitations come primarily from
the 5D kinetic energy terms, with a small contribution from the Higgs
interaction (which  gives mass to the zero-mode fields):
$$
m_n^2 \ = \ \frac{n^2}{R^2} \ + \ m_{SM}^2
$$
Since the compactification radius $R$ is of order of several
hundred GeV$^{-1}$, this means that the masses at a given KK level are
almost degenerate;
KK number conservation thus implies that the first
level excitations of light SM particles are stable.   
This degeneracy is, however, lifted by loop corrections \cite{cms1}.
The consequences of going beyond tree level can be thought of as
being twofold. First, there are radiative corrections due to the
fields propagating along the fifth dimensions (called bulk terms in 
\cite{cms1}). These corrections are well defined and finite, due
to $1/m_n^2$ suppression for heavier KK modes.
For the fermionic fields they are zero, while for the gauge fields they
are actually negative, and of order $\alpha/R$.
Second, loop effects induce boundary terms localized
on the fixed points of the $S_1/Z_2$ orbifold \cite{cms1,ggh}. 
The coefficients of these
terms depend on the fundamental theory at the Plank scale, and they are
unknown in the low energy regime; moreover, the contributions to these terms
coming from one loop 
corrections in the bulk are logarithmically divergent. Thus, 
it is necessary
to introduce a cutoff scale $\Lambda$; it is also necessary to specify
an {\it ansatz} for the definition of unknown coefficients. We shall follow
the choices made in \cite{cms1,cms2}; we refer the reader to these references
for more details, and here we just summarize the results. 

After taking into account the boundary terms contributions, the mass hierarchy
between the first level KK excitations is as follows. The heaviest particle
is the $g^*_1$, which acquires a positive 20-to-30\% correction to its mass
(depending on the choice of $\Lambda$). The next to heaviest particles are
the excitations of the SM quarks, for which the mass correction is in the 
20\% range. There is a small splitting between the mass of the $q^{\bullet}$
and $q^{\circ}$ quarks, due to differences in the electroweak interactions 
of the two fields. Since excitations of the top quark do 
not play a big role in our analysis, we can neglect the SM masses of the
quarks involved. The rest of the first level KK excitations, arranged in 
order of decreasing mass, are the heavy gauge bosons $W^*$ and $Z^*$, the 
$L^{\bullet}$ excitations of the lepton and neutrino fields and the
$l^{\circ}$ excitations of the same fields. The corrections to the masses
of these particles are below 10\%. Finally, the lightest KK excitation
is the $\gamma^*$, whose mass does not change almost at all from tree level. 

Due to the fact that the corrections to 
neutral U(1) and SU(2) gauge fields $\delta m_{B_n}$ and 
$\delta m_{A_n}$ are different, the mixing angle between these fields
will not be the SM Weinberg angle anymore. In the limit when
$\delta m_{A_n}^2 - \delta m_{B_n}^2 \gg m_W^2$, which generally holds
for values of the compactification scale greater than 200 GeV, the n$th$ level
mixing angle $\theta_{W_n}$ will actually be very close to zero \cite{cms1}. 
This means that the $\gamma^*$ will be almost a pure $B$ field; also, the
fermions which are singlets under SU(2) (the $q^{\circ}$ and $l^{\circ}$)
will decouple almost completely from the SU(2) fields. The interaction
between the electroweak gauge boson excitations and fermionic fields
will be given by:
$$
{\cal L}_{q - A_n} \ = \ - \ \bar{Q} \ e \left\{ \not{X_n}
\left[ Q \frac{\hbox{cos}\theta_{W_n}}{\hbox{cos}\theta_W} 
- \tau_3  \frac{\hbox{sin}(\theta_W - \theta_{W_n})}
{\hbox{sin}2\theta_W} \right] - 
\not{Z_n}
\left[ Q \frac{\hbox{sin}\theta_{W_n}}{\hbox{cos}\theta_W} 
- \tau_3  \frac{\hbox{cos}(\theta_W - \theta_{W_n})}
{\hbox{sin}2\theta_W} \right]
\right.
$$ 
$$
\left.
\ + \ \frac{g_2}{\sqrt{2}}\left[ 
\begin{array}{cc}
0 & \not{W^+} \\
\not{W^-} & 0
\end{array}
\right]
\right\} P_L\  Q^{\bullet}_n
- \ \ \bar{q}\ eQ \left( 
\not{X_n} \frac{\hbox{cos}\theta_{W_n}}{\hbox{cos}\theta_W}  -
\not{Z_n}
\frac{\hbox{sin}\theta_{W_n}}{\hbox{cos}\theta_W}  
 \right)
P_R \ q^{\circ}_n \ +\ \hbox{h.c.}
$$

These couplings, together with the mass hierarchy discussed above, dictates
the following decay pattern for the excitations of quarks and gluons
produced at a hadron collider (see also \cite{cms2}). The $g$ excitations
will decay equally through the $g^*_1 \rightarrow q \bar{q}^{\bullet}_1 $ and 
 $g^*_1 \rightarrow  q \bar{q}^{\circ}_1$ channels
(and the conjugate ones), where $q$ can be any quark
excludig the top, which is too heavy to be kinematically allowed. 
The singlet quark excitations $q^{\circ}_1$ will decay directly to the LKP:
$q^{\circ}_1 \rightarrow q \gamma^*$, since its coupling to the $Z^*_1$ boson
is suppressed. The decay of the doublet quark excitations $q^{\bullet}$ will
proceed mostly through a three stages chain:
\beq{chain} 
q^{\bullet}_1\ \rightarrow\ q\ Z^*_1\ \rightarrow 
q\ l\ l^{\bullet}_1\ \rightarrow\ q\  l\ l\ \gamma^* \ ,
\ \ \hbox{Br.} \sim 33\%, \ \hbox{and}
\eeq
$$ q^{\bullet}_1\ \rightarrow\ q\ W^*_1\ \rightarrow 
q\ l' \ l^{\bullet}_1\
 \rightarrow\ q\  l'\ l \ \gamma^* \ ,
\ \ \hbox{Br.} \sim 65\% $$
where $l$ and $l'$ can be either a lepton or a neutrino. The branching 
ratios (Br.'s) for the intermediate $Z^*_1$ and $W^*_1$ decays to leptons
are all approximatively equal to 1/6 (decays to quarks are not kinematically
allowed). A small portion of the $ q^{\bullet}_1$ decays (about 2\%) also
takes place directly to the LKP: $q^{\bullet}_1 \rightarrow q \gamma^*$.

In absence of other interactions, the LKP in UED is stable. Then, the
only signal of KK excitations production at a hadron collider will 
be missing energy in conjunction with soft leptons or jets radiated 
in the course of the decays of these excitations to the LKP. The 
phenomenological implications of this model have
been studied in \cite{cms2}. In this paper we aim to study the implications
of having the LKP decay into a KK graviton and SM photon. Then, the
experimental signal will be a striking two photon event with high $p_T$, plus 
large missing
energy and soft jets or leptons.  

There are a number of models which allow for the gravity mediated decay 
of KK excitations \cite{acd,rizzo,rigolin}\footnote{Note, also,
that the one-loop boundary terms may 
induce gravity-mediated LKP decays even in the
absence of new physics. We plan to study this possibility in a further paper.}. 
As in \cite{mmn}, for illustration
we will consider the specific case of a fat brane scenario \cite{rigolin}.
In this model, the 4+1 dimensional space in which the UED fields live
is a `fat' brane in the 4+N dimensional bulk in which gravity propagates. 
The compactification scale of the bulk dimensions $R_b$ is  of order 
eV$^{-1}$. Thus, in this model, the UED fields propagate a short way in the
fifth dimension (the width of the brane $\pi R$), while gravity propagates all
the way up tp $2 \pi R_b$. The couplings of the KK matter excitations
with gravity are proportional with the overlap of their wave functions along
the fifth dimension. The  extra momentum along the $y$ direction resulting
from KK number violation is absorbed by the brane.  The large density of
states for the KK gravitons in the fifth dimension 
(the splitting between adjacent
levels is of order eV) makes up for the smallness of the 
gravitational coupling, allowing the decay width of the matter KK excitations
through this mechanism to be phenomenologically relevant (i.e.,
they decay within the detector). More
details, as well as numerical
results for these gravity mediated decay widths can be found in \cite{mmn}.

\section{Collider Signals}

In this section we will discuss the possibility of discovery of UED 
KK excitations at the Tevatron and  LHC. We start by assuming that 
the gravitational decay widths of the first level KK excitations
are much smaller than the widths of the decays allowed by the mass splittings
among these particles. This can happen for example in the fat brane
scenario, if the number of extra dimensions in which gravity propagates
is $N = 6$. Then, the only role of KK number violating
gravity interaction is to mediate the decay of the $\gamma^*$ obtained
as a final result of the decays of gluon and quark excitations produced 
in the collision.    
\footnote{In this analysis we neglect KK particles produced 
through weak interactions,
since the cross-section is small.} 

Since the quarks and leptons radiated during the decay to $\gamma^*$ 
are soft, the experimental signal for the production of a pair of
KK excitations will be two photons plus a large amount of missing energy
(taken away by the KK gravitons). The main  
backgrounds for this signal are multijet, direct photon, $W + \gamma$,
$W$ + jets, $Z \rightarrow e e$ and  $Z \rightarrow \tau \tau \rightarrow e e$
events with misidentified photons and/or mismeasured $\not{E}_T$.
These backgrounds can be eliminated by using cuts on the transverse
momentum of the photons and the missing energy. 

\begin{figure}[t!] 
\centerline{
   \includegraphics[height=3.in]{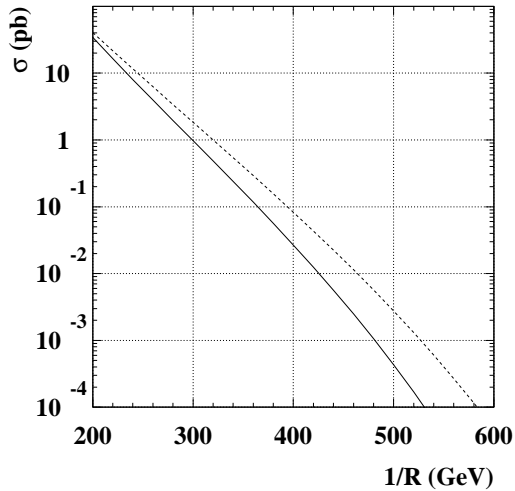}
   }
\caption{
Cross-sections for $\gamma \gamma X {\not E_T}$ signal coming
from universal extra dimensions at Tevatron Run I (solid line) and 
Run II (dashed line). The kinematic cuts applied are described in the text.}
\label{nu_mass}
\end{figure}
 
 The cross-sections for the $\gamma \gamma X {\not E_T}$ signal
coming from KK excitations production at Tevatron
are presented in Fig. 1. On the horizontal axis is the compactification
scale $1/R$. The solid line is the prediction for the Tevatron
Run I ($\sqrt{s} = 1.8$ TeV), while the dashed line is the prediction
for the Tevatron Run II ($\sqrt{s} = 2$ TeV). Here $N$ is taken to be 6.
These results also include cuts on photon
$p_T$ and ${\not E_T}$ as described below.

Searches for new physics in the $\gamma \gamma X {\not E_T}$ channel
at Tevatron have been performed by the D0 and CDF collaborations
\cite{2gD0,2gCDF}. For example, imposing the kinematic cuts on the
photon transverse momenta: $p_T^{\gamma_1} > 20\ \hbox{GeV},
p_T^{\gamma_2} > 12\ \hbox{GeV}$, and $ {\not E_T} > 25$ GeV, the DO
experiment finds two events (with an integrated luminosity of 106 pb$^{-1}$). 
The SM prediction from the above mentioned background is $2.3 \pm 0.9$
events. The solid line in Fig. 1 is the prediction coming from 
the UED model, with the kinematic cuts included.
In order to set precise lower bounds on the allowed
compactification scale $1/R$, a more detailed analysis, which includes
simulation of particles' interactions with the detector, is needed. However,
assuming a detector efficiency close to 1, the 
numbers above indicate that the 95\% CL upper
limit on the cross section is about 0.05 pb. This means that the Tevatron
Run I has excluded the UED model with gravity mediated decay of the 
LKP for scales up to about 380 GeV. 
  
 For the Tevatron Run II, the following cuts have been proposed in 
\cite{2grunII}: $p_T^{\gamma_1}, p_T^{\gamma_2} > 20$ GeV, 
$ {\not E_T} > 50$ GeV. The estimated SM background is $0.6 \pm 0.12$ fb 
(the large drop compared with Run I is mainly due to
 the increase in the $ {\not E_T}$
cut). Again, let's assume that the detector efficiency is 1.
Then, the 5$\sigma$ discovery cross section with these kinematic cuts 
is about 4.5 fb with 2 fb$^{-1}$
of integrated luminosity (Run IIA), 
or 1.2 fb with 15 fb$^{-1}$ integrated luminosity (Run IIB).
Then, UED
extra dimensions will be discovered at Run IIA if the compactification 
scale is smaller than about 490 GeV (520 GeV in Run IIB). 
On the other hand, assuming that the signal observed matches
the predicted background, exclusion of the
model at 95\% CL can be achieved for $1/R < 510$ GeV at Run IIA, or
$1/R < 540$ GeV at Run IIB.

\begin{figure}[t!] 
\centerline{
   \includegraphics[height=3.in]{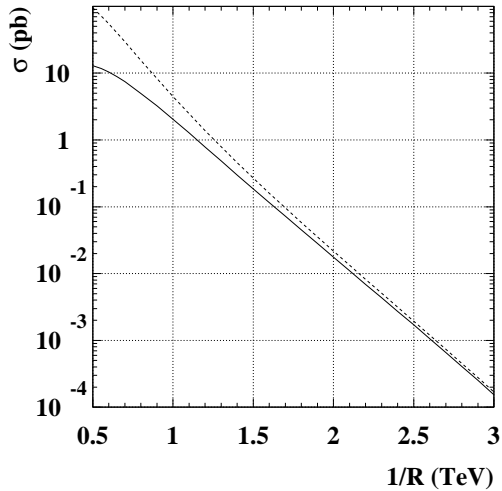}
   }
\caption{Cross-sections for $\gamma \gamma X {\not E_T}$ signal coming
from universal extra dimensions at LHC for $N=6$ (solid line) and 
$N=2$ (dashed line). The kinematic cuts applied are described in the text. }
\label{lhc}
\end{figure}

Fig. 2 contains the cross-sections for $\gamma \gamma X {\not E_T}$ 
production through KK excitations at the LHC. The solid line corresponds
to the case when there are $N = 6$ extra dimensions in which gravity
can propagate, while the dashed line is the result for $N = 2$.  The 
kinematic cuts on the momenta of the two hard photons are 
$p_T^{\gamma_1}, p_T^{\gamma_2} > 200$ GeV, 
$ {\not E_T} > 200$ GeV. The SM background with these cuts is estimated in 
\cite{2gtata} to be roughly 0.05 fb. The $5\sigma$ discovery cross-section
with 100 fb$^{-1}$ of integrated luminosity is then about 0.15 fb. We see
that the LHC can probe the compactification scale in this scenario up
to 3 TeV.

The analysis presented so far applies to the case
when the widths of the gravity mediated
decays of the KK excitations are smaller than the widths of the decays of one 
KK excitation to another. With this condition, the results presented above
are independent of the exact magnitude of these widths, and therefore
on the exact parameters of the model. If the opposite
case holds (that is, if gravity mediated decays dominate), then the
experimental signal for the UED scenario will be two jets + missing energy.
This case has been studied in \cite{mmn}. If, on the other hand, the two
sets of decay widths are roughly of the same order of magnitude,
more complicate decay patterns can ensue. Hard leptons (from the decay of
$q^{\bullet}$) can also appear in the final state, along with a combination
of photons and jets. The exact branching ratios depend on the parameters
of the model.

Note that the magnitude of the decay widths of the KK excitations can
depend quite strongly on the exact way the model is defined. The widths of 
decays among same level KK excitations depend on the masses involved, 
therefore on the cutoff scale $\Lambda$ (although this dependence is
only logarithmic), and, maybe more importantly, 
on the assumptions made in fixing the unknown
coefficients of the boundary terms. The widths of the gravity mediated
decays depend, of course, on the exact mechanism which induces these decays.
But even for a specific mechanism, let's take the fat brane scenario as
an example, they will depend on parameters like the number of dimensions
in which gravity propagates ($N$), or on the fundamental Planck scale 
$M_D$. As a consequence, an analysis of the interplay between
gravity and mass splitting effects in the decays of first level
KK excitations is bound to be quite model dependent.

In the results presented below we will use the framework described in
\cite{cms1,cms2} for the evaluation of the one loop mass corections to
the first level KK excitations. The cutoff scale $\Lambda$ is given by
$\Lambda R = 20$, and  the coupling constants are evaluated at the 
compactification scale. For the gravity sector we take $M_D = 5$ TeV, and
present results for $N=2$ and $N=6$. The widths of the gravity 
mediated decays are evaluated in accordance with the formulas given
in \cite{mmn}.

\begin{figure}[t!] 
\centerline{
   \includegraphics[height=3.in]{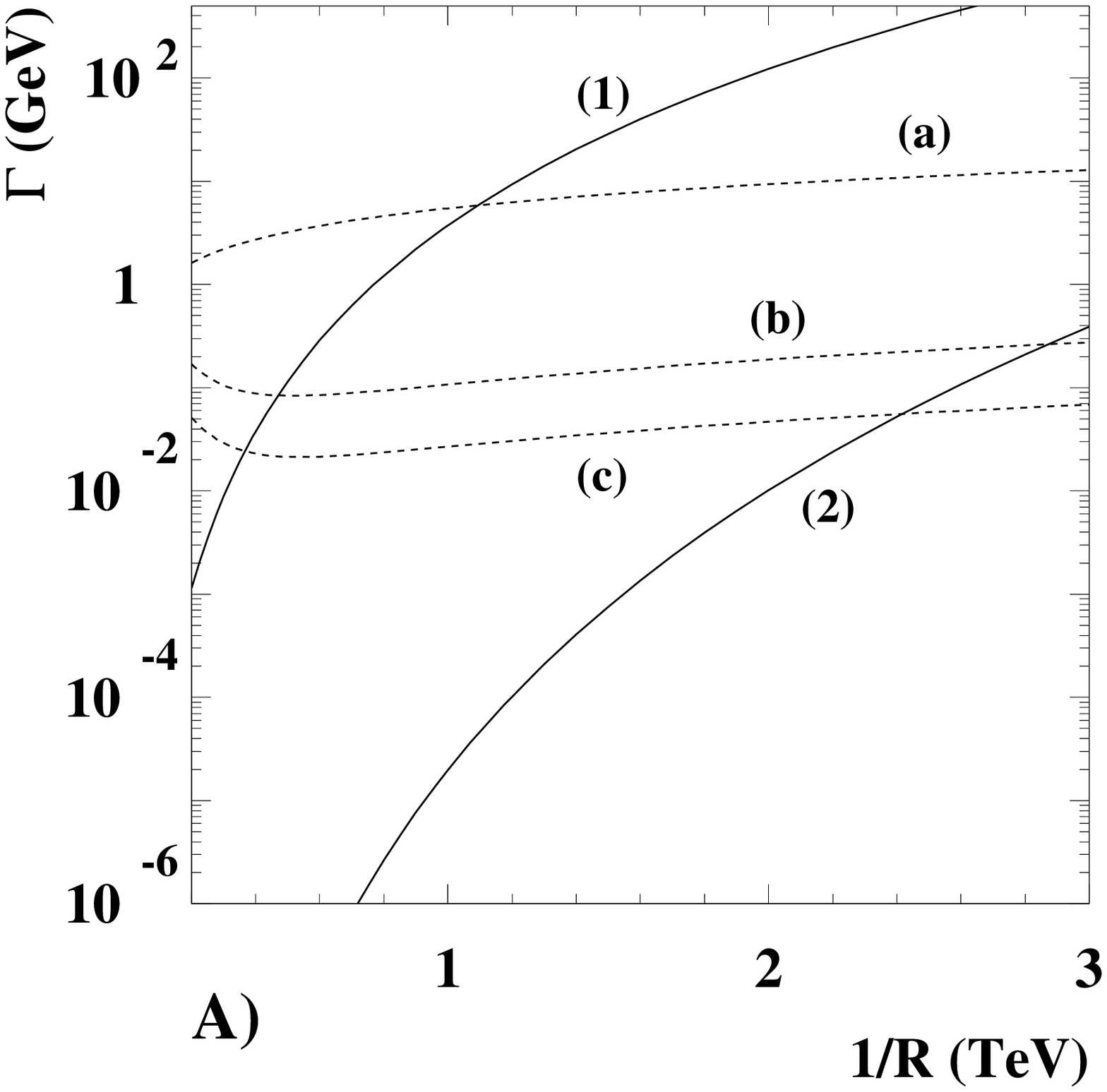}
   \includegraphics[height=3.in]{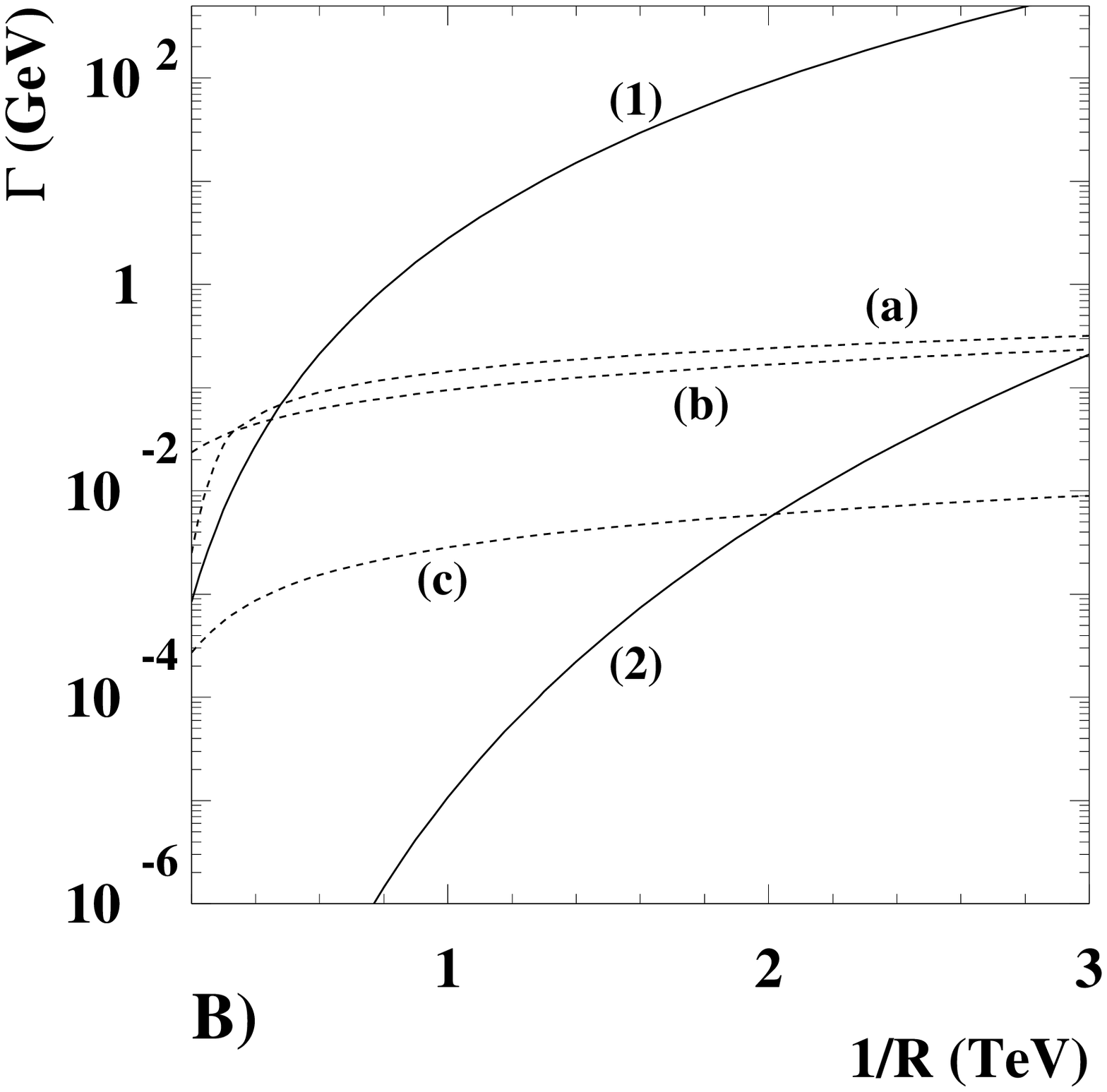}
   }
\caption{Decay widths for the first level KK excitations of gauge bosons (left)
and fermions (right). The solid lines correspond to gravity mediated decays,
with $N=2$ (1) and $N=6$ (2). The dashed lines correspond to decays allowed
by mass splittings: 
$g^* \rightarrow q \bar{q}^{\bullet} ,   q \bar{q}^{\circ} + \hbox{h.c.}$ 
(a), $W^* \rightarrow l \bar{\nu}^{\bullet}, \nu \bar{l}^{\bullet}
+ \hbox{h.c.}$ (b), $Z^* \rightarrow l \bar{l}^{\bullet}  + \hbox{h.c.}$
(c) (left), and 
$ q^{\bullet} \rightarrow q' W^* , q Z^*$ (a), 
$ q^{\circ} \rightarrow q \gamma^*$ (b),
$ l^{\bullet} \rightarrow l \gamma^*$ (c) (right).
}
\label{widths}
\end{figure}

\begin{figure}[b!] 
\centerline{
   \includegraphics[height=3.in]{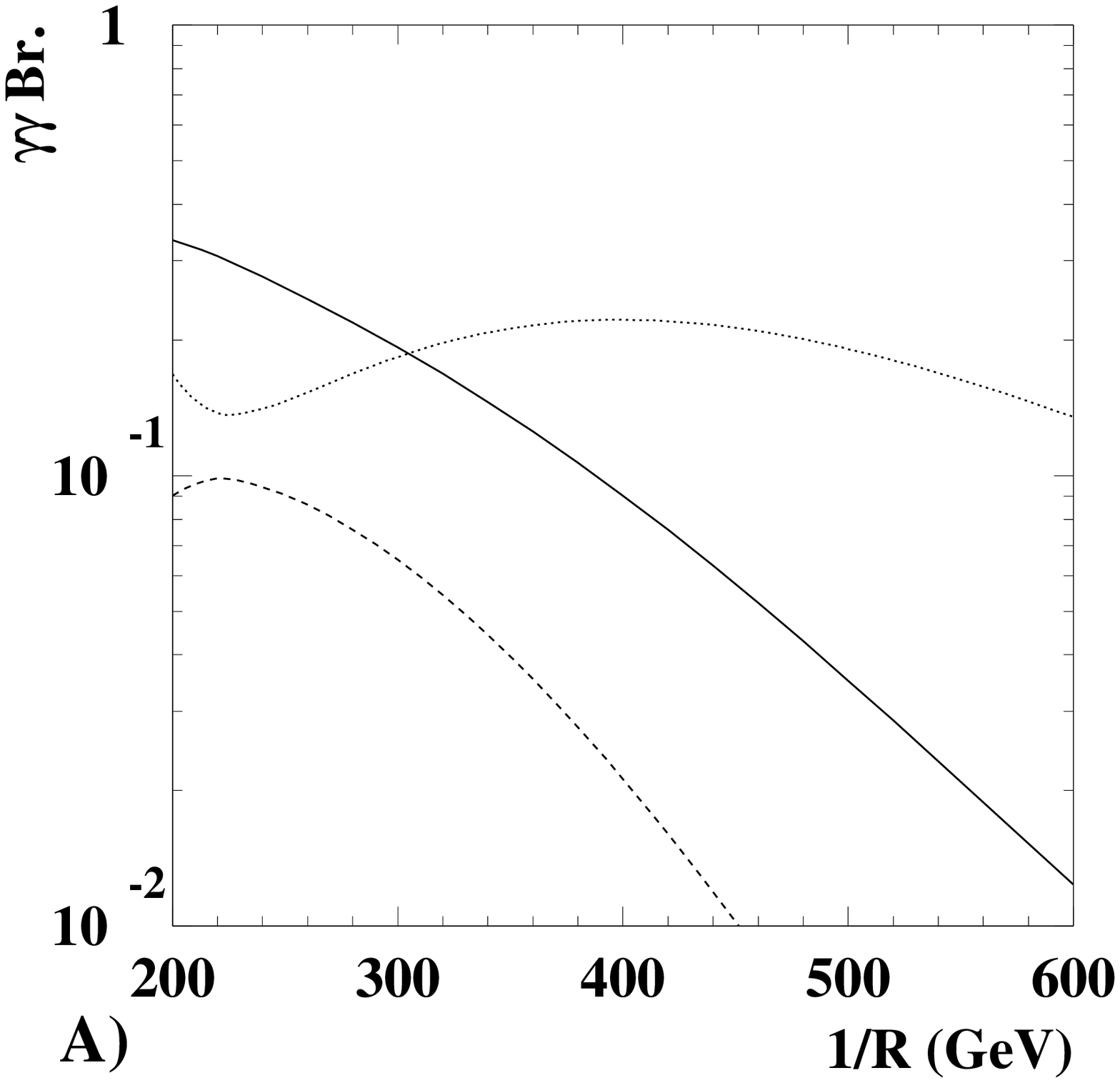}
   \includegraphics[height=3.in]{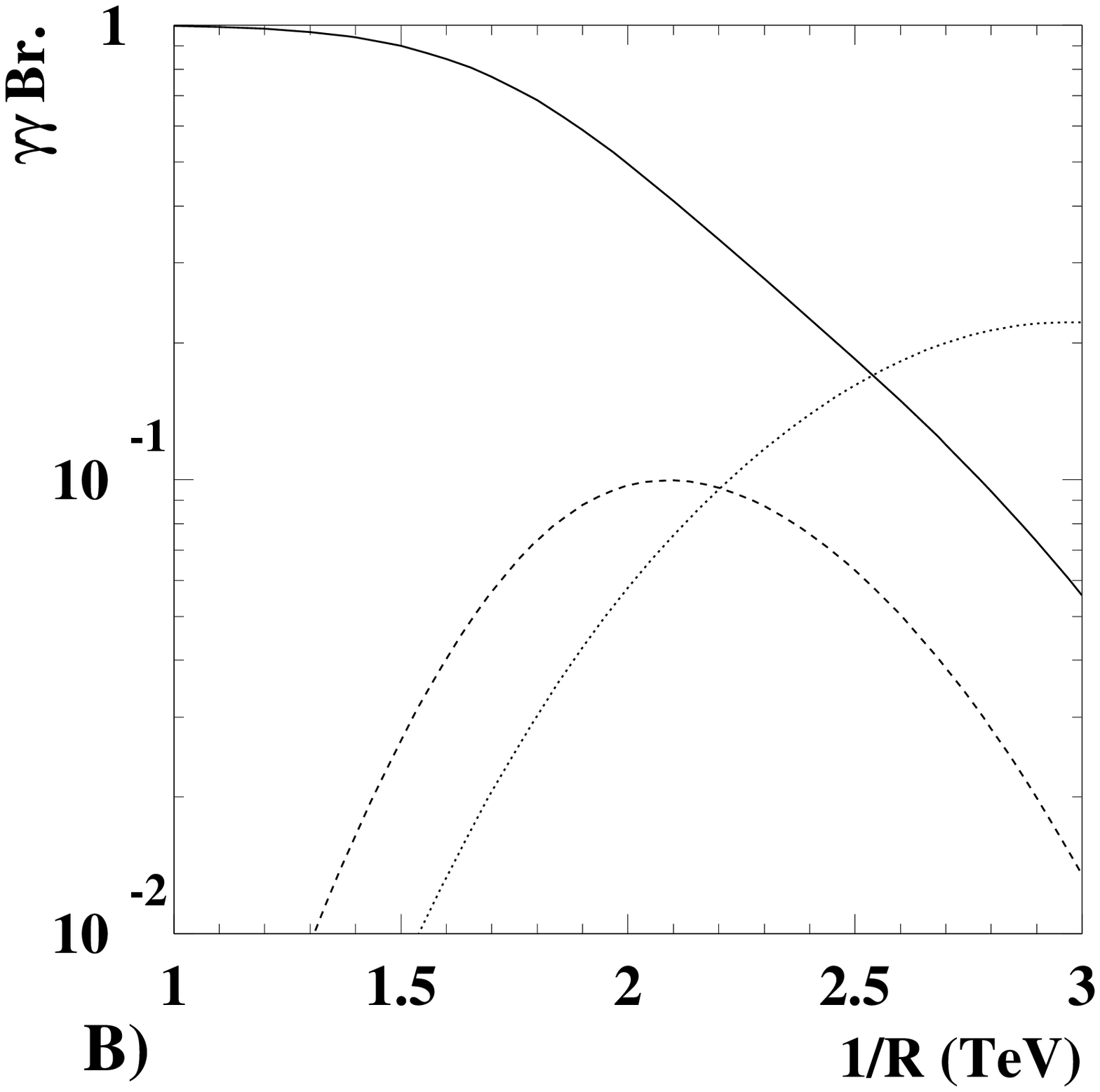}
   }
\caption{Branching ratios to final states: $\gamma \gamma$ (solid line),
jet + $\gamma$ (dotted line) and lepton ($e$ or $\mu$) + 
$\gamma$ (dashed line) at the Tevatron Run II (left) for $N=2$, and
LHC (right) for $N=6$. }
\label{bratio}
\end{figure}

Fig. 3 (A and B) 
contains the values of the decay widths of first level KK excitations. 
The decay widths of gauge bosons are presented in 3A; the solid lines
corrrespond to the gravity mediated decay widths for $N=2$ (1) and 
$N=6$ (2), while the dashed lines correspond to the widths 
of the decays due to mass splittings and gauge interactions: 
$g^* \rightarrow q \bar{q}^{\bullet} ,   q \bar{q}^{\circ} + \hbox{h.c.}$ 
(a), $W^* \rightarrow l \bar{\nu}^{\bullet}, \nu \bar{l}^{\bullet}
+ \hbox{h.c.}$ (b), and $Z^* \rightarrow l \bar{l}^{\bullet}  + \hbox{h.c.}$
(c).
Fig. 3B contains the decay widths of the fermions; again the solid lines
corrrespond to the gravity mediated decay widths for $N=2$ (1) and 
$N=6$ (2), while the dashed lines correspond to the widths of the decays
$ q^{\bullet} \rightarrow q' W^* , q Z^*$ (a), 
$ q^{\circ} \rightarrow q \gamma^*$ (b) (this is the result for up-type
quarks; the width for down-type quarks is four times smaller), and
$ l^{\bullet} \rightarrow l \gamma^*$ (c) ($ l^{\circ}$ is not produced
in the decay of the quark and gluon excitations). These results
can provide an estimate of what will be the main decay mode of the 
KK particles, and therefore what will be the experimental signal; for
example, we see that if $N=2$, gravity mediated decay will dominate, 
so we can expect two jets + $\not{E_T}$ in the final state; while
if $N=6$, the dominant signal for the UED model will be two photons 
+ $\not{E_T}$ for values of compactification scale up to about 2.5 TeV. 

In Fig. 4 we give the actual branching ratios for the $\gamma \gamma$
final state from the production of KK excitations. Note that these
results not include only information about the relative decay
widths, but also information about the relative ratios of 
$q^{\circ}$ quarks, $q^{\bullet}$ quarks and $g^*$
produced in the collision. (This is relevant since the $q^{\circ}$ quark,
which decays directly to the $\gamma^*$, has a higher branching ratio
to a final state containing a hard photon that the $q^{\bullet}$ or $g^*$,
which have to decay through a cascade.) 
Fig. 4A contains the $\gamma \gamma$ branching ratio for $N=2$ at low
compatification scale (this branching ratio is 1 for $N=6$), while
Fig. 4B shows the $\gamma \gamma$ branching ratio for $N=6$ at higher
compatification scale (the branching ratio being 0 for $N=2$ here).
For completeness we also include the branching ratios to 
jet$+ \gamma X \not{E_T}$ and $l \gamma X \not{E_T}$ final states, 
where $l$ is a hard $e$ or $\mu$ coming from the
gravity mediated decay of its first KK excitation, and $X$ stands for the 
soft particles radiated during the decay chain.

\section{Conclusions}

Large universal extra dimensions models have exciting implications for 
the phenomenology of future colliders. In this paper we have studied a
model where the signal for UED is an excess of two photon events 
(plus missing ${\not E_T}$) at hadron colliders. This final state arises
naturally in the context when: a) mass splitting between the 
first level KK excitations allows the 
the gluon and quark excitations 
produced in the initial collision to  cascade decay to the LKP (which is 
the $\gamma^*$); and b) gravity mediates the KK number violating
decay of the LKP into a hard photon, which shows up in the detector,
and a graviton, which is not observed. The condition for this chain
of events to take place is that the widths for mass splitting mediated
decays of the KK excitations be larger than the 
widths of the gravity mediated decays (while
the latter ones are sufficiently large that the $\gamma^*$ decay
happens in the detector). We discuss a specific realization 
of this model, where the one loop masses of KK particles are computed
in the framework used by Cheng, Matchev and Schmaltz in \cite{cms1,cms2},
and gravity mediated decays take place in the fat brane scenario 
described in \cite{mmn}.

 For cases when all the gluon and quark excitations decay to the LKP
first, Tevatron Run I data sets a 380 GeV lower limit on the
compactification scale for the UED. 
Run II will be able to probe for extra dimensions to  500 GeV with 
2 pb$^{-1}$ and 
540 GeV with 15 pb$^{-1}$, while  the LHC reach is about 3 TeV.
Limits in this channel 
will be correspondingly weakened when gravity mediated decays
of $g^*$ and $q^*$ also play some role. If these decays dominate, the search
for UED excitations should take place in the two jets plus 
missing ${\not E_T}$ channel, as analyzed in \cite{mmn}.

\vspace{1.cm}
{\Large \bf Acknowledgments}
\vspace{0.5cm}

 We thank M. Coca and R. Culbertson
for useful discussions concerning experimental limits.
This work was supported in
part by the U.S. Department of Energy Grant Numbers
DE-FG03-98ER41076 and DE-FG02-01ER45684.
C.M. and S.N. gratefully acknowledge  support from the Fermilab Summer
Visitor Program
and warm hospitality from the Fermilab Theory Group during the
completion of this work.


\end{document}